\documentclass[journal, 12pt, draftclsnofoot, onecolumn, web]{IEEEtran}
\usepackage{cite}
\usepackage{stfloats}
\usepackage{algorithm}
\usepackage{graphicx}
\usepackage{wrapfig}
\usepackage{textcomp}
\usepackage{amsmath}
\usepackage{amsfonts}
\usepackage{amssymb}
\usepackage{mathtools}
\usepackage{lipsum}
\usepackage{cuted}
\usepackage{comment}
\usepackage{xcolor}
\usepackage{algorithm}
\usepackage{algorithmicx}
\usepackage{algpseudocode}
\usepackage{caption}
\usepackage{subcaption}

\def\BibTeX{{\rm B\kern-.05em{\sc i\kern-.025em b}\kern-.08em
    T\kern-.1667em\lower.7ex\hbox{E}\kern-.125emX}}
\begin{document}
\title{A High Capacity Preamble Sequence for Random Access in Beyond 5G Networks: Design and Analysis}
\author{Sagar Pawar, Lokesh Bommisetty and T.G.Venkatesh

\thanks{Sagar Pawar, Lokesh Bommisetty and T.G Venkatesh are with the Electrical Engineering Department, Indian Institute of Technology Madras, Chennai, INDIA 600036}

}

\maketitle

\begin{abstract}
 The widely used Zadoff-Chu sequence (ZC sequence) for random access preamble in 5G has limitations in terms of the total number of preambles generated, forcing the reuse of preambles. Hence, the probability of collision of preambles of UEs increase, resulting in the failure of random access procedure. To truly qualify beyond 5G networks as green technology, the preamble capacity should be increased without sacrificing energy efficiency. In this paper, we propose a new candidate preamble sequence called $mALL$ sequence using the concept of cover sequences to achieve higher preamble capacity without degrading the power efficiency and hence minimizing device's carbon footprint. We compare the performance of $mALL$ sequence with Zadoff-Chu sequence and other sequences in the literature, such as $mZC$ and $aZC$ sequences. We evaluate the performance of the preamble sequences in terms of periodic correlation, detection probability and the effect of diversity combining. Also, this paper explores the Peak to Average Power Ratio (PAPR) and Cubic Metric(CM) for these sequences, as these are essential parameters to evaluate energy efficiency. We show that the preamble capacity of the proposed $mALL$ sequence is $10^{4}$ times higher than that of legacy ZC sequence without any deterioration in the detection performance.
%This paper presents the comparison between Zadoff-Chu, $m$-sequence and Alltop sequence and their detection performance in RA procedure for different number of receive antennas. 
%We further use the concept of cover sequences to get different combinations of Zadoff-Chu, $m$ and Alltop sequences for increased preamble capacity. Going beyond, we propose a new candidate sequence of $mAll$ with higher preamble capacity compared to the sequences proposed in the literature. The performance of Random Access Preambles in terms of periodic correlation, false-alarm probability and detection probability are analyzed for different preamble sequences. Also, this paper explores the Peak to Average Power Ratio (PAPR) and Cubic Metric(CM) for these sequences, as these are essential parameters to be considered during transmission.  
\end{abstract}

 
\section{Introduction}
\label{sec:introduction}
\IEEEPARstart{T}he deployment of 5G NR is a step forward for providing Enhanced Mobile Broadband (eMBB), Ultra Reliable Low Latency Communications (URLLC) and Massive Machine Type Communications(mMTC). Enhanced Mobile Broadband (eMBB) communications have requirements of data rates up to 10 to 20 Gbps with high mobility support of User Equipment (UE) \cite{campos2017understanding}. Ultra Reliable Low Latency Communications (URLLC) require robust connectivity and very low end to end latency of 5 ms\cite{campos2017understanding}. Massive Machine Type Communications(mMTC) aims for density of One million $(10^{6})$connections per square kilometer for IoT devices with low power requirements\cite{campos2017understanding}. Today, the internet and telecommunications is responsible for $4-6\%$ of total global power consumption. With the 5G and beyond 5G upgrades in the emerging markets, it is anticipated that the telecommunication sector triples their power consumption \cite{energyreport}. Hence, investigation and quantification of the energy efficiency of the 5G communication devices, signal transmitters and waveforms is a matter of importance  \cite{8654701}.  

%3GPP, in its Release 16, has defined two frequency ranges for operations which are FR1 band (Sub-6 GHz) and FR2 (millimetre wave) \cite{TS38.101}. The Sub-6 GHz band is already in use in LTE, which gives us a larger coverage area with low directionality. The use of mmWave in 5G NR is a key point in achieving higher throughput and supporting higher mobility speeds for UE. Nevertheless, the use of mmWave comes with its merits and demerits. mmWave communication allows higher sub-carrier spacing which in turn support higher bandwidths. The highly directional nature of mmWave results in a lesser number of multipath propagation, which requires simpler and faster estimators. However, these advantages come with a cost of higher path losses \cite{tataria2021standardization} as distance increases and the need for line of sight communication for best-case scenarios.

For a UE to establish an uplink connection, Random Access (RA) procedure is initiated by the UE\cite{TS38.321} after satisfying a few conditions. RA initiation requires the downlink connection to be established, where timing synchronisation between gNB and UE is essential. The downlink synchronisation is done with the help of Primary Synchronisation Signal (PSS) and Secondary Synchronisation Signals (SSS), using which the UE acquires information about Physical Cell Identity (PCI), Cell Specific Reference Signal (CSR) and resource allocations. Using this information, UE initiates RA procedure to establish an uplink connection with the gNB. 

The main component for the initiation of RA procedure is the preamble. The preamble is a fixed length sequence that is primarily used for uplink synchronisation. In LTE and 5G NR, ZC sequences are used as preamble sequences, but the length of the sequence restricts the number of different sequences that can be generated. The UE chooses one of the preamble from the available set of preambles randomly and initiates the RA procedure for uplink synchronisation and transmission. RA procedure of a UE is successful if the preamble chosen by it is not transmitted in the same resource units by any other UE, otherwise its a failure due to preamble collision \cite{TS38.321}. Especially for mMTC type communications, where there is a very high density of connection requests, the probability of collision of preambles becomes higher if ZC sequence is used, resulting in frequent failure of RA procedure \cite{TS38.321}. Hence, the energy requirements in the cell for random access increases rapidly with the increasing device density.

The motivation to explore preamble sequences other than ZC sequences is that, the use of ZC sequence limits the number of unambiguous sequences that can be generated. Exploring other sequences which can provide more distinct preambles can reduce the need for reuse of sequences and thus reducing the probability of collision and the energy requirements. In this context, several alternative sequences have been considered in the literature  \cite{pitaval2020overcoming}.

%The limited number of available preambles using ZC sequence create a need for other sequences to be studied that can provide more number of preambles which in result will reduce the reuse of a preamble and hence reduce the probability of collision. Several sequences have been proposed for such use in the literature \cite{pitaval2020overcoming} \cite{pitaval2018overcoming}.

Aim of this paper is to propose a new preamble sequence and to carry out a detailed performance study of the different preamble sequences proposed in the literature for the random access procedure in 5G. The major contributions of our paper are as follows.
\begin{itemize}
    %\item \textcolor{red}{\textbf{List down the contributions of the paper. eg: detection algorithm, sequences we used, metrics analysed etc. as separate points}}
    %\item We calculate the threshold for the proposed sequences, in order to maintain the false alarm probability $P(false) \leq 0.1\%$.
    \item We propose a new candidate preamble sequence called $mALL$ and show that $mALL$ has higher preamble capacity than other sequences proposed in the literature.
    \item We carry out a detailed performance study of the preamble sequences proposed in the literature for the random access procedure in 5G. 
    \item We evaluate the performance of the preamble using metrics that include periodic correlation, false detection, mis-detection. 
    \item We also explore other metrics related to preamble sequences such as PAPR and Cubic-Metric to better understand the energy requirements and inherent properties of such sequences.
    \item We observe the effect of number of antennas on detection performance of different preamble sequences, which provides us an insight on the effect of equal gain diversity combining on probability of detection of the preamble.
\end{itemize}

%\textcolor{red}{Aim of the paper is to}In this paper we look into the periodic correlation properties of different sequences proposed as candidates for preamble, their detection performance. We propose a new sequence with increase preamble capacity. We also explore other metrics related to the preamble sequences such as PAPR and cubic-metric to better understand the behaviour of such sequences.

%\textcolor{red}{\textbf{Write what we do in this paper in brief}}

Rest of the paper is organised as follows. In Section \ref{sec:background} we discuss the background of preamble generation and some important definitions. In Section \ref{sec:relatedwork}, we present the related work. We discuss the concept of cover sequences and compute the preamble capacity of existing cover sequences in Section \ref{otherpreamble}. We propose a new candidate sequence for preamble generation in Section \ref{proposedmall}. In Section \ref{preambledetection}, we lay out the algorithm used for preamble detection. In Section \ref{sec:simulations} we present the simulation results comparing the Detection Probabilities, PAPR, Cubic Metric and effect of Carrier Frequency Offset for the different sequences which are being considered as preamble candidates. Finally in Section \ref{conclusion}, we provide our conclusions regarding the features of preambles considered in the study.

\section{Background}
\label{sec:background}
The preambles used for the random access procedure in 5G are generated on the principle of orthogonality, where the set of preamble sequences are orthogonal to each other making it easy to detect at the receiver. Constant Amplitude Zero Auto Correlation (CAZAC) Sequences are considered to be suitable candidates for preamble generation. These sequences maintain Constant Amplitude so that the Peak to Average Power Ratio (PAPR) is low, which is required to maintain the linearity of Power Amplifiers. CAZAC sequences have another property, i.e. any two different sequences in a CAZAC sequence set are orthogonal to each other. A type of CAZAC sequences, called ZC sequences have been used in LTE and 5G NR for preamble generation. The length of a preamble sequence, denoted by $L_{RA}$, is based on the numerology defined by 3GPP technical specification \cite{TS38.211}. Before getting into the details of the preamble sequences, we introduce a few notations and definitions, that are used in the paper as follows.   

\subsection{Some Definitions}\label{sec: definitions}
\subsubsection{Cyclic Shift}\label{cyclicshift}
Let $x_\mu[n]$ be a sequence of length $L_{RA}$ with root index $\mu$, then the cyclic shifted $x_\mu^{n_0}[n]$ is defined as:

\begin{equation}
    x_\mu^{n_{0}}[n] = x_\mu[(n+n_{0}) \; mod \; L_{RA}] 
    \label{x_c}
\end{equation}
where $n_0$ is the number of samples by which $x^{n_{0}}[n]$ is cyclically shifted from $x[n]$. Here, $mod$ represents the modulo operation i.e., $a\; mod\;b$ gives the value of the reminder when $b$ divides $a$.
\subsubsection{Periodic Correlation}\label{periodiccorrelation}
Periodic correlation of any two sequences $x[n]$ and $y[n]$ of length $L$ is given by the following equation.

\begin{equation}
    R_{x,y}[\tau] = \sum_{n=0}^{L_{RA}-1} x[n] y^{*}[(n+\tau) \; mod \; L_{RA})] 
    \label{R_xy}
\end{equation}

\noindent
where $y^*[n]$ represents the conjugation of sequence $y[n]$

\subsubsection{Zero / Low Correlation Zone (ZCZ) / (LCZ)}\label{zczlcz}
The ZCZ / LCZ is defined as the region where the value of $|R_{x,y}|$ is zero/very low as compared to the peak value of $|R_{x,y}|$.

\subsubsection{Probability of Detection / Misdetection} 
\label{P-missP-detect}
$P(detection)$ is defined as the conditional probability of correct detection of the preamble when the signal is present. $P(miss)$ is complementary of $P(detection)$. Standard (3GPP, TS 38.104 \cite{TS38.104}) dictates that $P(detection)$ should be equal to or exceed $99\%$ \cite{TS38.104}.

\subsubsection{Probability of False Alarm $(\;P(false)\;)$} \label{pfalse}
$P(false)$ is defined as the conditional probability of erroneous detection of the preamble when input is noise only. $P(false)$ should be less than $0.1\%$ \cite{TS38.104}.

\subsubsection{Preamble Capacity} \label{pcapacity}
Preamble Capacity can be defined as the total number of preambles from which a set of 64 preambles is generated for each cell. As the 5G NR aims for mMTC, there will be a need to increase the number of preambles per cell. We have a limited number of preamble sequences depending on the length of sequence and the allowed cyclic shift between preamble sequences. The method of generating the preamble sequences is illustrated in TS 38.211 (\cite{TS38.211}, Sec. 6.3.3.1). In case of ZC sequences of length $L_{RA}$, we have $L_{RA}-1$ different root sequences and for each root sequence we have $ \left \lfloor{\frac{L_{RA}}{N_{CS}}} \right \rfloor$ different cyclically shifted sequences. Therefore preamble capacity for ZC sequence is given as:
\begin{equation}
    PrCapacity^{ZC} = (L_{RA}-1).\left \lfloor{\frac{L_{RA}}{N_{CS}}} \right \rfloor
    \label{prcapacityzc}
\end{equation}
From \eqref{prcapacityzc} we clearly see the dependence of preamble capacity on $L_{RA}$, the length of preamble sequence, and $N_{CS}$, the minimum cyclic shift between two sequences. As $N_{CS}$ increases, the number of preambles available decreases. This may lead to the reuse of sequences among the UEs in the cell when the UEs randomly choose the available preambles during the RA procedure. This results in high probability of collision. If the number of unique preambles available is large, then the reuse of sequences will decrease. So, there is a need for the study of other sequences or the combination of some sequences such that higher preamble capacities are obtained. These sequences can then be used instead of ZC sequence to cater the needs of mMTC. 

\section{Related Work}
\label{sec:relatedwork}
  ZC sequences and its properties have been extensively studied by Frank \textit{et al.} \cite{frank1962phase} and Chu \textit{et al.} \cite{chu1972polyphase}. The restricted length of preamble sequences, limit the number of different preambles generated for a ZC sequence as discussed in Section \ref{pcapacity}. In addition to the low preamble capacity, ZC sequences are found to be sensitive to frequency shift by Pitaval \textit{et al.} \cite{pitaval2020overcoming}. Schreiber \textit{et al.} \cite{schreiber20185g} proposed a new candidate sequence for preamble generation called m-Sequence. The m-Sequences show better tolerance to frequency shifts. To increase the number of preambles available, several techniques have been proposed in the literature. Aggregation techniques are explored by Mostafa \textit{et al.} \cite{mostafa2018aggregate} where the transmitted preamble is a weighted addition of two ZC sequences. Arana \textit{et al.}  \cite{arana2017random} designed a new preamble by multiplying shifted versions of two different ZC sequences. The combination of ZC sequences with other CAZAC or near-CAZAC sequences has been suggested by Pitaval \textit{et al.} \cite{pitaval2020overcoming}

Vukovic \textit{et al.} \cite{vukovic2015impact} has systematically analysed the impact of AWGN channel on RACH throughput. A detection technique has been proposed by Pham \textit{et al.} \cite{pham2019proposed} for long sequences i.e., $L_{RA}=839$ for rejecting false peaks. A preamble design and detection method for satellite random access is discussed by Zhen \textit{et al.} \cite{zhen2020preamble}. S. Kim \textit{et al.} \cite{kim2012delay}, T. Kim \cite{kim2017enhanced} proposed more sophisticated detection schemes that make use of post-processing and reconstructions of received signal in order to achieve better detection but with increased complexity.

Although a number of preamble sequences have been proposed and studied in the literature, the effect of diversity combining on the detection performance has not been explored in detail for these preamble sequences. Furthermore, Cubic Metric (CM), which is an important parameter, has not been studied for these sequences. To meet the demand of increasing devices especially in mMTC, the preamble capacity should be as high as to serve $10^6$ devices per $km^2$. To fill in the said gaps in the literature, this paper aims to find the detection performance using diversity combining of the proposed sequences and their CM measurements, which can help to identify better overall sequences as candidates for preambles.

Unique contributions of our paper are as follows

\begin{itemize}
    
 \item We propose a new candidate preamble sequence called $mALL$ by combining the $m-sequence$ and $Alltop$ sequences. Further, we show that $mALL$ has higher preamble capacity than other sequences proposed in the literature.
 
  \item Unlike the existing works in literature \cite{pitaval2020overcoming}\cite{schreiber20185g}, in addition to the PAPR performance, we also study Cubic-Metric performance of different preamble sequences for better understanding of the inherent properties of such sequences.
  
  \item Ours is the only work to present the effect of equal gain diversity combining on the detection performance of the preamble sequences considered in \cite{pitaval2020overcoming} \cite{schreiber20185g} and for the proposed $mALL$ sequence.
  
  \item We capture the effect of carrier frequency offset on the correlation properties of the proposed preamble sequences $\{mALL\}$ and compare it with the legacy ZC sequences.  
  
  \end{itemize}

%\section{Random Access procedure: Preamble Principles}\label{sec:preamble principles}

%A brief review of properties and definitions {para}

\section{Existing Preamble Designs and their Preamble Capacity}\label{otherpreamble}

\subsection{Concept of cover sequences}\label{coverseq}
From \eqref{prcapacityzc} we note that, we have a fixed number of preamble sequences that can be generated based on $L_{RA}$ and $N_{CS}$. Let us denote this set of preambles available in ZC sequences as $S^{ZC}$. 
Consider another CAZAC sequence such as $m-sequence$ \cite{schreiber20185g} and $Alltop$ \cite{alltop1980complex} sequence of the same length $L_{RA}$. We call it a cover sequence and use it to generate more number of unique preambles. We denote the cover sequence as  $c[n]$ . Let $N_{c}$ be the number of sequences that can be generated such that they are all orthogonal to $c[n]$ and non-ambiguous to one another. Here non-ambiguity implies that the cross-correlation between any two sequences in $S^c$ is very low compared to the total signal energy, so that they can be identified and detected easily. We define this set of $N_{c}$ cover sequences as $S^{c}$. 
\begin{equation}
\begin{split}
    S^{c}=\{ c_{1}[n],c_{2}[n],c_{3}[n], \dots ,c_{N_{c}}[n] \}
    \label{coverset}
\end{split}
\end{equation}

We define a set of orthogonal signals as $s_{\mu}^{ZC}$ such that for a fixed $\mu$ and different $v$, the generated sequences are orthogonal to each other. Therefore we have
\begin{equation}
    \begin{split}
        s_{\mu}^{ZC}= \left \{ x_{\mu}^{0},x_{\mu}^{1},x_{\mu}^{2},...,x_{\mu}^{ \left \lfloor{\frac{L_{RA}}{N_{CS}}} \right \rfloor - 1 }\right \}
        \label{eq:11}
    \end{split}
\end{equation}

Union of $s_{\mu}^{ZC}$ for all values of $\mu$ will generate a super set $S^{ZC}$ such that $S^{ZC} = \bigcup_{\mu=1}^{L_{RA-1}} s_{\mu}^{ZC}$.
       
New combination sequences are generated when each sequence from $S^{c}$ multiplies with each sequence from $S^{ZC}$ element-wise as shown below. 
%\begin{equation}
 %   \begin{split}
  %  y_{l,\mu,v}[n] = c_{l}[n].*x_{\mu}^{v}[n], \quad & l=1,2,\dots,N_{c} \\
   %                                     & \mu=1,2,\dots,L_{RA}-1 \\
    %                    & v=0,1,\dots,\lfloor{\frac{L_{RA}}{N_{CS}}}\rfloor \\
     %                   & n=0,1,\dots, L_{RA}-1
      %                  \label{covereq}
    %\end{split}
%\end{equation}

\begin{equation}
    \begin{split}
    y_{l}[n] = & c_{l}[n].\ast S^{ZC} \\
     = & c_{l}[n].\ast\left \{ s_{1}^{ZC}, s_{2}^{ZC},.....s_{L_{RA}-1}^{ZC} \right \}  \quad l=1,2,...,N_{c}
    \label{covereq}
    \end{split}
\end{equation}
Here in \eqref{covereq}, the notation '$.\ast$' represents element-wise multiplication.
We define a set $S_{l}^{ZC}$, which is a set of sequences obtained from \eqref{covereq} for a fixed value of $l$ which can be varied from $1$ to $N_{c}$. The union of all such $N_{c}$ sets will result  into a superset $S^{cZC}$: 
\begin{equation}
    S^{cZC}= \bigcup_{l=1}^{N_{c}} S_{l}^{ZC}
    \label{quasiset}
\end{equation}

Popovic in \cite{popovic2011quasi} defines such set in \eqref{quasiset} as Quasi-Orthogonal Set.  
\subsection{$m-sequence$ and combination with ZC sequence($mZC$)}\label{mseq}
$m-sequences$ are binary sequences generated by using Linear Feedback Shift Registers (LFSR)\cite{schreiber20185g} \cite{pitaval2020overcoming}. Generation of these sequences are defined by its primitive generator polynomials. A $m^{th}$ order polynomial is defined as follows.
\begin{equation}
    g(P)=g_{m}P^{m} + g_{m-1}P^{m-1} + \dots + g_{1}P^{1} + g_{0}
    \label{generatorpoly}
\end{equation}
Where $g_{m}$ is the polynomial coefficient which can take value of either $0$ or $1$. 

% \begin{figure}
%     \centering
%      \includegraphics[width=\linewidth]{m-seq-circuit.png}
%      \caption{$m-sequence$ generator using LFSR where P is a shift register, $g_{m}$ is the polynomial coefficient in Eq. \eqref{generatorpoly} and $\bigoplus$ represents X-OR operation }
%      \label{fig:m-seq-circuit}
% \end{figure}

The length of sequence is given by $N_{m}=2^{m}-1$. For example consider $m=7$, we have a length of sequence as $N_{m}=127$ and generator polynomial as $P^{7} + P^{1} + 1$. Using these polynomial weights, LFSRs will generate a periodic binary bit streams $x_{m}[n]$ of length 127.
%\textcolor{red}{CHANGE n0 to l}
We can obtain a set of orthogonal sequences $S^{m}$ from $x_{m}[n]$ by cyclic shifting the sequences.
\begin{equation}
    x_{m}^l[n] = x_{m}[(n+l) \; mod \; N_m] 
    \label{cyclicmseq}
\end{equation}
\begin{equation}
    \begin{split}
        S^{m} = \bigcup_{l=0}^{N_{m}-1} \{x_{m}^{l}[n]\}
        \label{setmseq}
    \end{split}
\end{equation}

% \begin{figure}
%     \centering
%      \includegraphics[width=\linewidth]{mZC_corr.png}
%      \caption{Periodic Correlation of $mZC$ sequences, where $l,\mu,v$ are varied. The absolute auto-correlation of $z_{l,\mu,v}$ in Eq. \eqref{eq:mZCseq} where $\{l,\mu,v \}$ = \{0,1,0\} is represented by blue line . Next four results represented in the legend are the cross-correlation results between two different sequences $z_{l,\mu,v}$ and $z_{l',\mu',v'}$with different values of $\{l,\mu,v \}$.   }
%      \label{fig:mZC}
% \end{figure}

The inherent issue with the m-sequences is that they are defined perfectly only for lengths of $2^{m}-1$ i.e., 3, 7, 15, 31, 63, 127, 255, 511 and so on. As defined by 3GPP technical specifications\cite{TS38.211}, the length for short preamble is defined as $L_{RA}=139$. The value of $m$ for which the length for $m-sequence$ is closest to $L_{RA}$ is $m=7$, corresponding to 127 length $m-sequence$. At the defined lengths the cyclic shifted versions of $m-sequences$ defined in \eqref{cyclicmseq} are orthogonal to each other. But the need for 139 length sequences as cover sequences raises the compatibility issue with $m-sequences$. To generate 139 length cover sequence, the first 12 samples of $m-sequence$ are appended to itself. Although the new sequences generated lose their orthogonality, they maintain non-ambiguity since their cross-correlation is very low compared to the total energy of signal. Hence, they behave similar to near-CAZAC sequences. 

As mentioned in \ref{coverseq}, we can use the $m-sequences$ as a cover sequence to obtain the new preamble sequence, $z_{l,\mu,v}$ and the Quasi-Orthogonal set $S^{mZC}$ \cite{pitaval2020overcoming} as given below:
\begin{equation}
    z_{l,\mu,v} [n]= \left (x_{m}^{l}[n] \right ).*x_{\mu}^{v}[n]
    \label{eq:mZCseq}
\end{equation}
\begin{equation}
    S^{mZC}=\bigcup_{l,\mu,v}\{z_{l,\mu,v}[n]\}
\end{equation}

To calculate the preamble capacity of $mZC$ sequence, we see that we can use the set $S^{ZC}$ as it is. Then we have $N_{m}$ number of sequences in set $S^{m}$. Therefore multiplying each sequences in $S^{m}$ with each sequence in $S^{ZC}$ we will have $N_{m}(L_{RA}-1)\left \lfloor{\frac{L_{RA}}{N_{CS}}} \right \rfloor$ sequences along with the original ZC set. Therefore we have 
\begin{equation}
    \begin{split}
    PrCapacity^{mZC}&= N_{m}(L_{RA}-1)\left \lfloor{\frac{L_{RA}}{N_{CS}}} \right \rfloor\\ &+ (L_{RA}-1)\left \lfloor{\frac{L_{RA}}{N_{CS}}} \right \rfloor \\
    & =(N_{m}+1)(L_{RA}-1)\left \lfloor{\frac{L_{RA}}{N_{CS}}} \right \rfloor
    \label{prcapMZC1}
    \end{split}
\end{equation}

The maximum value for $N_{m}$ can be attained using a cyclic shift of each sample which gives us $N_{m}^{max}=L_{RA}$. Substituting this value in \eqref{prcapMZC1}, we have

\begin{equation}
    \begin{split}
    PrCapacity^{mZC}= (L_{RA}^{2}-1)\left \lfloor{\frac{L_{RA}}{N_{CS}}} \right \rfloor
    \label{prcapMZC}
    \end{split}
\end{equation}

% In Fig. \ref{fig:mZC}, we have plotted the correlation properties of mZC sequences formed in \eqref{eq:mZCseq}. We change the parameters $n_0,\mu,v$ and check for different combination to verify that the periodic auto-correlation gives us a peak at centre, but cross-correlation yield very low values at all locations. This leads to unambiguous signal detection.

% \begin{figure}
%     \centering
%      \includegraphics[width=\linewidth]{aZC_corr.png}
%      \caption{Periodic Correlation of $aZC$ sequences. We represent the auto-correlation result in blue line, where the sequence $z_{l,\lambda,w,\mu,v}$ is generated from Eq. \eqref{alltopzseq} with $\{ l,\lambda,w,\mu,v \}$ = \{1,1,1,20,0\}. Other two results represented in the legend are cross-correlation results between two sequences $z_{l,\lambda,w,\mu,v}$ and $z_{l',\lambda',w',\mu',v'}$ with different values of $\{ l,\lambda,w,\mu,v \}$ }
%      \label{fig:aZC}
% \end{figure}

\subsection{$Alltop$ combined with ZC sequence($aZC$)}\label{Alltop}
Alltop sequences as defined in \cite{alltop1980complex} are cubic phase sequences. One form of such sequences are defined in \cite{pitaval2020overcoming}, \cite{heath2006quasi} as follows:
\begin{equation}
    \begin{split}
        a_{\lambda,w}[n] = \exp { \left \{-j2\pi\frac{(n+w)^{3}+\lambda n}{L_{RA}}\right\}}, &\quad 0\leq n \leq L_{RA}-1 \\
        & 0 \leq \lambda \leq L_{RA-1} \\
        & 0 \leq w \leq L_{RA-1}
        \label{alltopeq}
    \end{split}
\end{equation}
Some important properties of Alltop sequences are given below:
\begin{itemize}
    \item $a_{\lambda,w}[n]$ is a cyclic shifted version of the sequence $a_{\lambda}[n]$.
    \item Two sequences $a_{\lambda,w}[n]$ and $a_{\lambda',w}[n]$ are orthogonal to each other for $\lambda \neq \lambda '$.
    \item Whereas, $a_{\lambda,w}[n]$ have very low cross correlation with $a_{\lambda,w'}[n]$ and is equal to $\sqrt{L_{RA}}$.
\end{itemize}
Applying the sequence in \eqref{alltopeq} as a cover to ZC sequence introduces ambiguity in resulting sequences. Pitaval \MakeLowercase{\textit{et al.}} in \cite{pitaval2020overcoming} discusses this in details and proposes a new cover sequence to be used as:

\begin{equation}
    \begin{split}
        z_{l,\lambda,w,\mu,v} = &\left (a_{\lambda,w}[n] \right )^{l}.*x_{\mu}^{v}[n], \quad  0 \leq l \leq L_{RA}-1
        \label{alltopzseq}
    \end{split}
\end{equation}

Preamble capacity of $aZC$ sequence is shown to be \cite{pitaval2020overcoming} 
\begin{equation}
    PrCapacity^{aZC} =\left ( L_{RA}^{2}-1 \right ) \left \lfloor{\frac{L_{RA}}{N_{CS}}} \right \rfloor
    \label{prcapAZC}
\end{equation}

% In Fig.\ref{fig:aZC} we have plotted the periodic correlation results of $aZC$ sequences for different combinations of values of $l,\lambda,w,\mu,v$. For same sequence, we observe that the auto-correlation results in a peak at centre and  for other cyclic-shifts values we have LCZ.

% Keeping $\lambda,w,\mu,v$ the same and varying $l$, we observe that we obtain correlation values which are not constant for different cyclic-shifts, but have an upper bound of $2\sqrt{L_{RA}}$ \cite{pitaval2020overcoming}. The values we obtain are very low compared to the peak obtained in auto-correlation and can be considered as LCZ.
%  Keeping the $l,\lambda,\mu,v$ the same and varying the cyclic shift of Alltop sequence, for $w=20$, we see a zero value at $20^{th}$ location, which implies orthogonality between two sequences and for  other values of cyclic-shift, it remains in LCZ. We cannot use all different combinations of the values $l,\lambda,w,\mu,v$ to generate preambles\cite{pitaval2020overcoming}, because ambiguity is present in some of the combinations. Hence the limit given in \eqref{prcapAZC}.

\section{Proposed $mALL$ preamble sequence} \label{proposedmall}
%\subsection{Combination of m and Alltop sequences (mALL)}

In Section. \ref{mseq} and \ref{Alltop}, we saw that both $mZC$ and $aZC$ sequences have the same $PrCapacity$ given by \eqref{prcapMZC} and \eqref{prcapAZC}. It is observed that for $aZC$ sequences, there can be multiple sequences which are ambiguous for an arbitrary choice of $l,\lambda,w,\mu,v$, reason being  $\lambda,w$ and $\mu$ parameters modify the phases of sequences simultaneously\cite{pitaval2020overcoming}. So we need to be careful in the selection of such sequences. In worst case scenario that can be obtained by substituting $\left \lfloor{\frac{L_{RA}}{N_{CS}}}\right \rfloor=1$ in \eqref{prcapMZC} and \eqref{prcapAZC}, we have $PrCapactity$ in order of $10^5$. However the requirement for mMTC is in order of $10^{6}$ per $km^{2}$\cite{campos2017understanding}. 

For Alltop sequences defined in \eqref{alltopeq} and \eqref{alltopzseq}, we find that for a fixed value of $l$ and $\lambda$, the $w$ parameter makes a cyclic shifted version of the sequence. Thus taking correlation of two Alltop sequences $\left (a_{\lambda,w}[n] \right )^{l}$ and $\left (a_{\lambda,w^{'}}[n] \right )^{l}$ will give us a high correlation peak, which limits the possible number of unique sequences that can be generated based on $l$ and $\lambda$ only. To make the correlation peak of Alltop sequences independent of $w$, we make use of m-sequences as cover sequences.

% \textcolor{red}{As $m$-sequences do not have any phase change associated with them, we expect to have no simultaneous phase change in the combined sequence of $m$ and $Alltop$.  For this reason, we propose a new sequence by combining the $m$ and $Alltop$ sequences, wherein we expect to avoid the ambiguity related to the simultaneous phase change in the sequences.}

We propose a new combination sequence called $mALL$ sequence by using m-sequence as a cover sequence for Alltop sequence. Therefore, from equations \eqref{alltopeq} and \eqref{cyclicmseq}, we define $mALL$ sequence to be,

\begin{equation}
    \begin{split}
         z_{l,\lambda,w,t} = \left (a_{\lambda,w}[n] \right )^{l}.*x_{m}^{t}[n] \\
         0\leq l,\lambda,w,t \leq L_{RA}-1 \\
         \label{eq:mALL}
    \end{split}
\end{equation}

\begin{figure}
    \centering
     \includegraphics[width=0.6\linewidth]{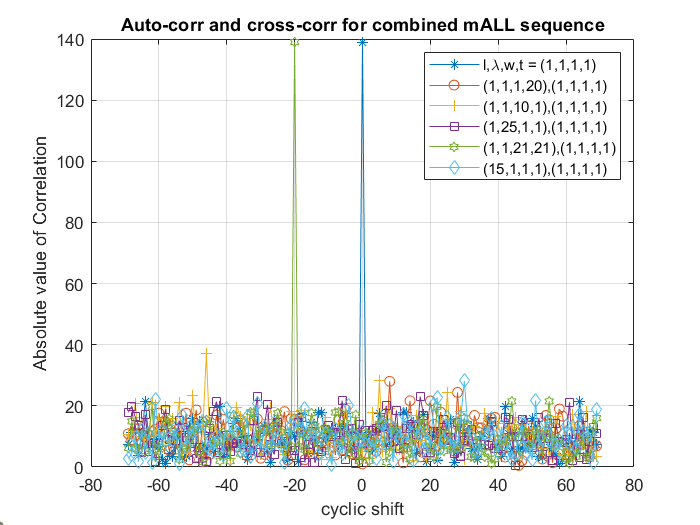}
     \caption{Periodic Correlation of $mALL$ sequences.We represent the auto-correlation result in blue line, where the sequence $z_{l,\lambda,w,t}$ is generated from Eq. \eqref{eq:mALL} with $\{ l,\lambda,w,t \}$ = \{1,1,1,1\}. Other  results represented in the legend are cross-correlation results between $z_{1,1,1,1}$ and $z_{l,\lambda,w,t}$ for arbitrary choice of $\{l,\lambda,w,t\}$ values.}
     \label{fig:mALL}
     \vspace{-4mm}
\end{figure}

We have generated the preamble sequences defined by \eqref{eq:mALL} in a simulation setup using MATLAB and have explored the auto-correlation and cross-correlation properties of this sequence. We run the simulation extensively for all possible combinations of the parameters, $l,\lambda,w,t$. Fig. \ref{fig:mALL} shows the correlation results for different combinations of $l, \lambda\ w, t$. Our findings are presented below:

\begin{itemize}
    \item The auto-correlation peak has a value equal to the length of a sequence $L_{RA}=139$ with all other samples in LCZ.
    \item The average maximum cross-correlation value between two different sequences is 27.37 which is less than $3\sqrt{L_{RA}}$. 
    \item For a fixed $t$ and varying $l,\lambda,w$ we find that all sequences generated are non-ambiguous where their cross-correlation values are always in LCZ.
    \item For fixed values of $l,\lambda$ and $w$, different values of $t$ lead to ambiguity in the sequences resulting in high correlation peak.
\end{itemize}

 It is observed in Fig. \ref{fig:mALL}, that the peak for auto-correlation of $z_{1,1,1,1}$ occurs at zero cyclic shift value and has a value equal to length of sequence ($L_{RA}=139$). It is observed that there exist another sequence $z_{1,1,21,21}$ which also shows peak of value of 139. Such sequences are ambiguous and can result in mis-detections. To avoid such cases, we simply keep $t$ constant and vary only $l,\lambda,w$ to obtain a larger set.
For other cases of cross-correlation, it is observed that the values remain sufficiently lower, such that detection of sequences using the method described in Section. \ref{preambledetection} holds good.
Therefore, each sequence obtained by varying $l,\lambda,w$ and keeping $t$ constant, we will have a set of unambiguous sequences. The number of sequences that can be generated from the range of values taken by $l,\lambda,w$ in equation \eqref{eq:mALL} we have $L_{RA}^{3}$ unique sequences.

To further increase the capacity we include cyclic shift of each sequence in the set, depending on $zeroCorrelationZoneConfig$. For each combination of $l,\lambda,w$,with a fixed $t$, we include the cyclic shift parameter $v$ such that

\begin{equation}
    \begin{split}
         z_{l,\lambda,w,t}^{v} = z_{l,\lambda,w,t}[(n+C_{v}) \; mod \; L_{RA}] 
         \label{eq:mALLcs}
    \end{split}
\end{equation}

\noindent
 where $C_{v}$ is the cyclic shft defined in TS 38.211 (\cite{TS38.211}, Sec. 6.3.3.1). Let $ \{z_{l,\lambda,w,t}^{v}[n] \}$ be the set of all non-ambiguous sequences for a given $v$ obeying \eqref{eq:mALLcs}.
Let us define a set of such sequences as:
\begin{equation}
    \begin{split}
        s^{mALL}_v = \bigcup_{l,\lambda,w} \{z_{l,\lambda,w,t}^{v}[n] \}
        \label{eq:mALLsubset}
    \end{split}
\end{equation}
From \eqref{eq:mALLsubset}, we get $L_{RA}^3$ sequences for each value of $v$ and all possible combination of $l,\lambda,w$, with a fixed $t$ . Union of all such sets in \eqref{eq:mALLsubset} over $v$ will form a complete set of sequences as given below.

\begin{equation}
    \begin{split}
        S^{mALL} = \bigcup_{v=0}^{\left \lfloor{\frac{L_{RA}}{N_{CS}}} \right \rfloor-1} s^{mALL}_v
        \label{eq:mALLset}
    \end{split}
\end{equation}

Taking all combinations into account we can calculate PRACH capacity for the proposed sequence:
\begin{equation}
    \begin{split}
        PrCapacity^{mALL} & = L_{RA} \times L_{RA} \times L_{RA} \times \left \lfloor{\frac{L_{RA}}{N_{CS}}} \right \rfloor \\
        & = L_{RA}^{3} \left \lfloor{\frac{L_{RA}}{N_{CS}}} \right \rfloor \\
        \label{prcapMALL}
    \end{split}
\end{equation}
We see from \eqref{prcapMALL}, that the proposed $mALL$ sequence has a much larger $PrCapacity$ than the previously identified sequences namely $ZC$, $mZC$ and $aZC$.

%\section {System Model}\label{systemmodel}
%The $P(detection)$ and $P(miss)$ are calculated using the detection method described in Section \ref{preambledetection}.

\begin{algorithm}[!h]
\caption{\textbf{Preamble Detection Algorithm}} \label{algo1}
\algorithmicrequire $Received \;Signal \;s_{i}[n] \;for\; i^{th}\; antenna $ \\
\algorithmicrequire $Root \;sequences\; k_{\mu}[n] $ \\
\algorithmicrequire $Threshold\; \rightarrow \eta^{N_{Ant}},\;N_{CS}$ \\
\algorithmicensure $L_{RA}=139$
\begin{algorithmic} 
    \State \textbf{Start:} 
    \State $S_{i}[k] \;\leftarrow \; \mathcal{F}(s_{i}[n])$  \Comment{FFT of received signal}
    \State $K_{\mu}[k]  \; \leftarrow  \; \mathcal{F}(k_{\mu}[n])$ \Comment{FFT of Root Sequences} 
    \State $P_{\mu}  \; \leftarrow  \; 0$ \Comment{Initialise PDP matrix}
\For{$i^{th}\; antenna$} 
    \For{$each \;row \;(\mu) \;of\; K_{\mu}[k]$} 
        \State $R_{s_{i},k_{\mu}} \; \leftarrow  \; \mathcal{F}^{-1}(K_{\mu}^{*}[k].S_{i}[k])$  \\ \Comment{periodic correlation}
        \State $P_{\mu}^{i}  \; \leftarrow  \; |R_{s_{i},k_{\mu}}|^{2}$  \Comment{Calculate PDP}
    \EndFor
    \State $P_{\mu} \;  \leftarrow  \; P_{\mu}+P_{\mu}^{i}$  \Comment{Accumulation of PDP}
\EndFor 
\State $ \overline{P}_{\mu} \; \leftarrow \; mean \{ P_{\mu} \}$
\For{$each\; row \;\mu$}
    \For{$i=0;i<L_{RA};i=i+N_{CS}$} 
        \State $window  \; \leftarrow \;  P_{\mu}[i:i+N_{CS}-1]$\\ \Comment{Window of length $N_{CS}$}
        \If{$MAX\{ window \} \geq \eta^{N_{Ant}}  \overline{P}_{\mu}$}\\ \Comment{Compare Threshold}
            \State$DETECT \leftarrow 1$  \Comment{Decision for Detection}
        \Else
            \State $DETECT \leftarrow 0$ \Comment{Not Detected}
        \EndIf 
    \EndFor 
\EndFor
\end{algorithmic}
\label{detectionalgo}
\end{algorithm}
\section{Preamble Detection Algorithm} \label{preambledetection}
For each cell there are 64 \footnote{This is valid for $L_{RA}=139$. For $L_{RA}=839$ there are restricted sets which limit the number of available preamble for transmission} preambles available at UE for transmission. This set of 64 sequences are generated according to standard specifications \cite{TS38.211}. The gNB has the information about which root sequences are present in all 64 preamble from which UE transmits one of the preamble randomly. To calculate the periodic correlation we make use of FFT (fast fourier transform) for faster calculations. As we know the convolution of two sequences in time domain is equivalent to the operation of element wise multiplication of the FFT of sequences. Let the received sequence be $s[n]$ and the root sequences present at gNB be $k_{\mu}[n]$.

\begin{equation}
    \mathcal{F}(s[n])=S[k], \qquad \mathcal{F}(k_{\mu}[n])=K_{\mu}[k]
    \label{fft}
\end{equation}
where $\mathcal{F}$ represents the Discreet Fourier Transform operator.
\begin{equation}
    \begin{split}
    R_{s,k_{\mu}}[\tau] & = \sum_{n=0}^{L-1} s[n] k_{\mu}^{*}[(n+\tau) \; mod \; L)] \\
                  & = \mathcal{F}^{-1}(K_{\mu}^{*}[k].S[k])    
                  \label{corrfft}
    \end{split}
\end{equation}

The detection algorithm adopted by us based on \cite{pitaval2020overcoming}\cite{pham2019proposed} is given in Algorithm \ref{algo1}:
\begin{figure}
    \centering
    \includegraphics[width=0.6\linewidth]{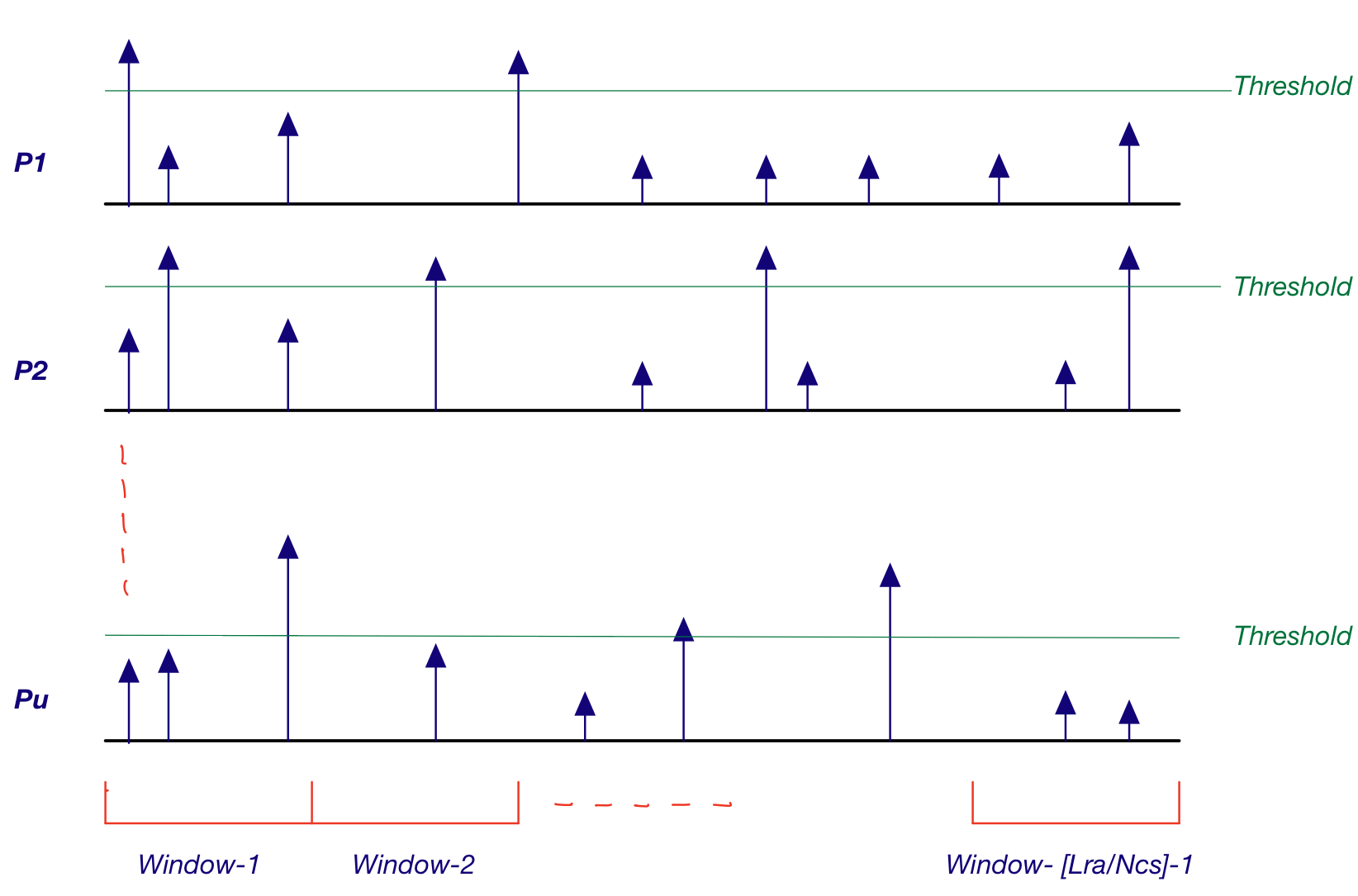}
    \caption{Illustration of detection algorithm}
    \label{fig:illus}
    \vspace{-4mm}
\end{figure}

\begin{itemize}
    \item First we calculate FFT of received signal $s_{i}[n]$ for $i^{th}$ antenna  and Root sequences $k_{\mu}[n]$
    \item Next we calculate the periodic correlation between $s_{i}[n]$ and $k_{\mu}[n]$ based on Eq. \eqref{corrfft}. This results into a matrix $R_{s_{i},k_{\mu}}$ of dimension $\mu \times L_{RA}$ where $1 \leq \mu \leq \left \lceil \frac{64}{\left \lfloor{\frac{L_{RA}}{N_{CS}}} \right \rfloor} \right \rceil$.
    \item Taking absolute squared value of $R_{s_{i},k_{\mu}}$, results into PDP matrix $P_{\mu}^i$. Here for each antenna we accumulate the $P_{\mu}^i$ matrix. The accumulated matrix is of dimension $\mu \times L_{RA}$.
    \item Fig. \ref{fig:illus} shows an example of a accumulated PDP where $P_{1}, P_{2},...,P_{\mu}$ correspond to the rows of $P_{\mu}$ matrix of length $L_{RA}$. The amplitudes denote the absolute squared values of the periodic correlation $R_{s,k_{\mu}}$. Each row is divided into $\left \lfloor{\frac{L_{RA}}{N_{CS}}} \right \rfloor$ number of windows of length $N_{CS}$. We declare the detection of preamble if the peak value in any window corresponding to any row is greater than the threshold $(shown\;by\;green\;lines)$.
    \item Then we calculate mean of $P_{\mu}$ obtained after accumulation as $\overline{P}_{\mu}$. 
    \item Next step is to divide each row of $P_{\mu}$ into $\left \lfloor{\frac{L_{RA}}{N_{CS}}} \right \rfloor$ different windows of length $N_{CS}$ as shown in Fig. \ref{fig:illus}.
    \item We then decide a threshold for each row as $\eta^{N_{Ant}}  \overline{P}_{\mu}$, shown as a green line in Fig. \ref{fig:illus}.
    \item If the peak value in any given window across the rows of $P_{\mu}$  is greater than  $\eta^{N_{Ant}}  \overline{P}_{\mu}$, we say that a preamble is detected. Based on the row and window number we decide which preamble out of 64 preambles was transmitted.
\end{itemize}

%\textcolor{red}{
%\begin{itemize}
 %   \item Calculate $FFTs$ of received sequence $s[n]$ and $k_{\mu}[n]$
  %  \item Using \eqref{corrfft} calculate the periodic correlation. Let $R_{s,k_{\mu}}[\tau]$ be the $\mu^{th}$ correlation results corresponding to each root sequence.
   % \item Calculate the Power Delay Profile (PDP) of each result, as:
    %\begin{equation}
     %   P_{\mu} = |R_{s,k_{\mu}}[\tau]|^{2}
      %  \label{pdp}
    %\end{equation}
    %\item Accumulate the obtained PDP in \eqref{pdp} from all antennas non-coherently using equal gain combining technique.See \ref{diversitycombining} %put reference
    %\item Calculate Threshold $\eta$ \ref{threshold} %put reference
   % for detection of preamble.
    %\item Divide the PDPs obtained from \eqref{pnorm} into a windows of length $N_{CS}$ defined by $zeroCorrelationZoneConfig$.
    %\item In each window for every sample check for the peak value. If the peak value is greater than $\eta$ then the corresponding preamble is detected.
%\end{itemize}
%}

\subsubsection{Diversity Combining} \label{diversitycombining}
Equal gain combining \cite{brennan2003linear}, is a linear diversity combining technique which is very simple to implement. It requires the received noisy signals from different antennas to be simply added before any processing.This technique can give us a significant SNR gain \cite{yue2019diversity}. In \cite{sun2014antenna} the use of equal gain combining for mmWave and MIMO application have also been discussed.

Let $N_{Ant}$ denote the number of receiver antennas at gNB. After equal gain combining at the receiver, the PDPs is given by:
\begin{equation}
    P_{\mu}^{N_{Ant}} = \sum_{i=1}^{N_{Ant}} P_{\mu}^{i}
    \label{eqgaincombine}
\end{equation}

\subsubsection{Threshold Measurement} \label{threshold}
Each of the PDP obtained from \eqref{eqgaincombine} is normalised by its average value $\overline{P}_{\mu}^{N_{Ant}}$.
\begin{equation}
    \overline{P}_{\mu}^{N_{Ant}} = \frac{1}{L_{RA}}\sum_{k=0}^{L_{RA}-1} P_{\mu}^{N_{Ant}}[k]
    \label{pdpavg}
\end{equation}
Therefore from \eqref{eqgaincombine} and \eqref{pdpavg} we have normalised PDP
\begin{equation}
    P_{\mu}^{Norm} = \frac{P_{\mu}^{N_{Ant}}}{\overline{P}_{\mu}^{N_{Ant}}}
    \label{pnorm}
\end{equation}

Compare the value obtained in \eqref{pnorm} with a decided value of $\eta$. If the value of $P_{\mu}^{Norm}$ is greater than $\eta$ then preamble is detected else it is not.

The value of $\eta$ is decided such that the condition for $P(false)$ defined in \ref{pfalse} is satisfied.
 
\section{Simulations and Results} \label{sec:simulations}
\subsection{{Simulation Methodology}} \label{method}
We make use of MATLAB tool for simulation purposes. We generate different sequences using Eq. \eqref{eq:mZCseq}, \eqref{alltopzseq} and \eqref{eq:mALLcs}. We generate a set of 64  preamble sequences set according to in TS 38.211 (\cite{TS38.211}, Sec. 6.3.3.1). We use the value for $zeroCorrelationZoneConfig$ as 11 which translates to $N_{CS}$ value of 23. The preamble set of 64 sequences is generated by varying the associated parameters for the sequences as  given in TABLE \ref{tab:64preambles}

\begin{table}[b]
     \centering
      \caption{Generation of $ZC$, $mZC$, $aZC$ and $mALL$ sequences where $L_{RA}=139$ and $N_{CS}=23$ }
     \label{tab:64preambles}
     \begin{tabular}{|c|c|c|} 
     \hline
     \textbf{Sequences} & \textbf{Variation of parameters $\{v={1\;to\;5},\; N_{CS}=23\}$ } &\textbf{Equation} \\
     \hline
     $ZC$ &  $\mu=\{1\;to\;11\}$ & Eq. \eqref{x_c}\\
     \hline
     $mZC$ &  $l=1, \; \mu=\{1\;to\;11\}$  & Eq. \eqref{eq:mZCseq}    \\ %1333080
     \hline
     $aZC$ & $l=1,\; \lambda=1,\; \mu=\{1\;to\;11\},\; w=1$ & Eq. \eqref{alltopzseq}  \\ %1333080
     \hline
     $mALL$ & $l=1 ,\; \lambda= \{1\;to\;11\},\;  w=1,\; t=1$ & Eq. \eqref{eq:mALLcs}\\ %185307711
     \hline
     \end{tabular}
 \end{table}

We use the AWGN channel for comparison of detection performance. For detection performance we use the method described in Section. \ref{preambledetection}. We run the simulation for $10^{5}$ times for each of the sequences, for every antenna to get $P(detect)$. 

\begin{figure}
    \centering
     \includegraphics[width=0.6\linewidth]{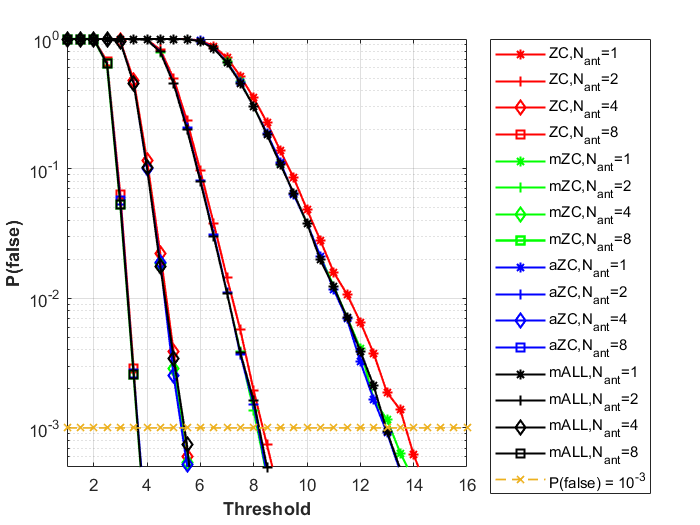}
     \caption{Estimating threshold for P(false) $\leq$ 0.1$\%$}
     \label{fig:Threshold}
     \vspace{-4mm}
 \end{figure}
 
\subsection{Threshold for False Alarm Probability} \label{threshold_calc}
Based on Section. \ref{threshold}, we determine the threshold to be set which satisfies the conditions for $P(false)$ given in Section. \ref{pfalse}. 
For simulation we follow the steps using Algorithm. \ref{detectionalgo}:
\begin{itemize}

    \item The received signal $s[n]$ in Algorithm. \ref{detectionalgo} is a complex AWGN noise.
    \item We correlate the received noise signal $s[n]$ with all root sequences $k_{\mu}[n]$.
    \item We decide the threshold value $\eta^{N_{Ant}}$ based on the number of antennas at the receiver.
    \item Based on  $\eta^{N_{Ant}}$, we make a decision for detection.
    \item We then have the probability of detection when input is AWGN noise for all sequences $\{ZC, mZC, aZC, mALL\}$ for each antenna configuration $\{N_{Ant}\}$.
\end{itemize}

In Fig. \ref{fig:Threshold}, we observe that as the Threshold increases the $P(false)$ decreases. It is found that to satisfy the condition of $P(false) \leq 0.1 \%$, we have Threshold defined in Table \ref{tab:Threshold}. We observe that there is a variation of $\eta$ for lower $N_{Ant}$ values but as $N_{Ant}$ increase, the $\eta$ is same for all four sequences.
\begin{table}[b]
\vspace{-4mm}
     \centering
     \caption{Threshold ($\eta$) satisfying $P(false)$ condition}
     \label{tab:Threshold}
     \begin{tabular}{|p{1.5cm}|p{1.25cm}|p{1.25cm}|p{1.25cm}| p{1.25cm}|} 
     \hline
     \textbf{Sequences} & \textbf{$\eta$ ($N_{Ant}$=1)} & \textbf{$\eta$ ($N_{Ant}$=2)} & \textbf{$\eta$ ($N_{Ant}$=4)} & \textbf{$\eta$ ($N_{Ant}$=8)} \\
     \hline
     $ZC$ & $13.7$ & $8.4$ & $5.4$ & $3.8$ \\
     \hline
     $mZC$ & $13.1$ & $8.2$ & $5.3$ & $3.8$  \\
     \hline
     $aZC$ & $13.0$ & $8$ & $5.3$ & $3.8$ \\
     \hline
     $mALL$ & $13.0$ & $8.4$ & $5.4$ & $3.8$ \\
     \hline
     \end{tabular}
 \end{table}

 \begin{figure}
    \centering
     \includegraphics[width=0.65\linewidth]{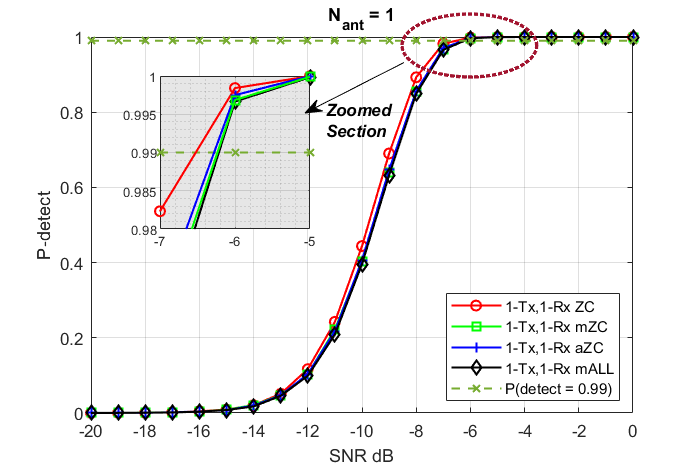}
     \caption{$P(detection)$ for $ZC$, $mZC$, $aZC$, $mALL$ for 1-TX,1-RX configuration}
     \label{fig:p-detect-single-ant}
     \vspace{-5mm}
 \end{figure}

\subsection{Detection Probability} \label{det-prob}
We use the configuration of single transmit and single receive antenna for this analysis. The parameter $zeroCorrelationZoneConfig$ is set to 11 that translates to $N_{CS}=23$. The $P(detection)$ is calculated using the detection method described in Section \ref{preambledetection} and values of $\eta$ taken from TABLE \ref{tab:Threshold}. In Fig. \ref{fig:p-detect-single-ant}, we observe that $ZC$, $mZC$, $aZC$ and $mALL$ sequences achieve $P(detect)$ of $99\%$ at -6.5 dB, -6.2 dB, -6.3 dB and -6.2 dB respectively. It is observed that the ZC sequence detection performance is better than rest of the sequences, and $mZC$ and $mALL$ sequences perform equally. Difference between the best performing ZC sequence and proposed sequence $mALL$ is only 0.3 dB, which is very low.

\begin{figure}[t]
    \centering
     \includegraphics[width=0.6\linewidth]{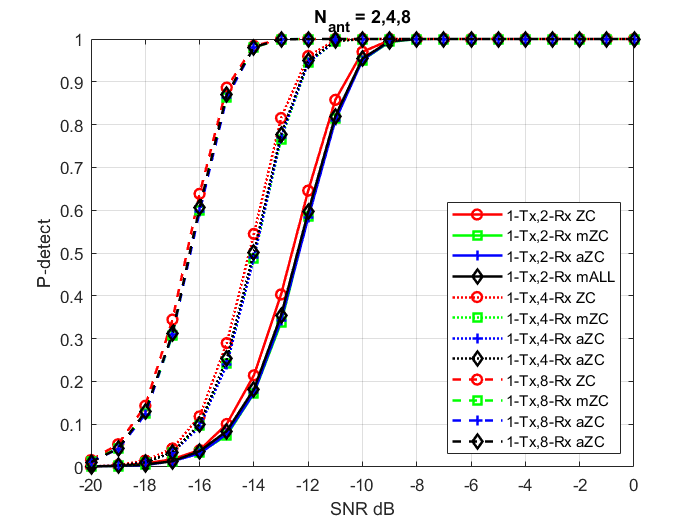}
     \caption{Comparison of $P(detection)$ for $ZC$, $mZC$, $aZC$, $mALL$ sequences varying SNR and $N_{Ant}$  }
     \label{fig:p-detect}
     \vspace{-4mm}
 \end{figure}
 
\subsection{Diversity Combining} \label{div-comb}
In Fig. \ref{fig:p-detect}, the detection performance has been observed in AWGN channel for a single user case, varying $N_{Ant}$ as 2,4 and 8. We wanted to study the properties of the combination sequences \{$mZC,aZC$ and $mALL$\} and compare their properties and detection performance with the $ZC$ sequence. In order to do that the preamble sequences are not repeated unlike the formats prescribed by 3GPP in \cite{TS38.211}. Threshold $\eta$ is taken from TABLE \ref{tab:Threshold}. We observe that there is no deviation in performance of detection with respect to varying SNR for all the four sequences considered namely $ZC$, $mZC$, $aZC$ and $mALL$. But varying $N_{Ant}$ has effect on the performance. All the four sequences perform in a similar manner for a given threshold calculated based on the number of antennas. However, as the number of antennas increase we find improvement in detection performance. This behaviour can be attributed to the diversity combining of antennas that results in more accurate detection.

In Fig. \ref{fig:p-detect}, for lower SNR in the range of -20 to -16 dB, the performance of the system for the cases of $N_{ant}=8$ and $N_{ant}=4$ is 3dB and 1dB better than that of the $N_{ant}=2$ case, respectively. When achieving the $P(detect)$ greater than $99\%$, $N_{ant}=8$ and $N_{ant}=4$ systems perform better than $N_{ant}=2$ system, by 4.6dB and 2.1dB, respectively. This is an significant improvement that we obtain using the diversity combining. Note, that the performance for legacy ZC sequence is always marginally better than that of other three sequences and the proposed $mALL$ sequence also performs marginally better than $mZC$ and $aZC$ sequences. It is also observed that as $N_{Ant}$ increases, the performance gap of $mZC$,$aZC$,$mALL$ with $ZC$ sequence reduces. 

Observations from Fig. \ref{fig:p-detect},  show that the detection performance of the proposed $mALL$ sequence is at par with ZC sequence and slightly better than $aZC$ and $mZC$ sequences. The advantage of proposed $mALL$ sequence is that, it provides a higher preamble capacity without the degradation of detection performance.

\subsection{PAPR and CUBIC METRIC (CM)} \label{cubic-metric}
PAPR and CM are the metrics which indicate the efficiency of Power Amplifier (PA) used to transmit the signals. Generally, higher the PAPR or CM, lower is the efficiency of the PA. PAPR gives us the measure of power backoff required. For a transmitted signal $x(t)$ PAPR is defined in \cite{rahmatallah2013peak} as:
\begin{equation}
   \begin{split}
       PAPR[x(t)] {dB}= 10 \log{\frac{max[x^{2}(t)]}{mean[x^{2}(t)]}}
       \label{papr-eq}
   \end{split}
\end{equation}

However, recent studies suggest that the CM is a better performance metric than the PAPR metric \cite{behravan2006some} \cite{zhu2014minimizing}\cite{huang2018reducing} as it considers the effect of 3rd order harmonics introduced due to the non-linearity in PA. In general, the amplifier voltage can be written as

\begin{equation}
    v_{0}(t) = g_{1}v_{i}(t) + g_{3}v_{i}^{3}(t) 
    \label{pa-gain}
\end{equation}
where $g_{1}$ and $g_{3}$ are linear and non-linear gains respectively. These gains are inherent to PAs and do not change with the type of signal. The cubic term in \eqref{pa-gain} introduces distortions in the transmitted signal, resulting in erroneous detection at receiver. Moreover, it also introduces third harmonics which interfere with adjacent channels. In order to suppress these distortion we need to suppress the non-linear gain ($g_{3}$). The CM for a transmitted signal $x(t)$ is given by \cite{rahmatallah2013peak}:

\begin{equation}
    \begin{split}
   CM[x(t)] = \frac{20\log{rms[x_{norm}^{3}(t)]}-1.52}{1.56} \\
   where \quad x_{norm}(t) = \frac{\mod{x(t)}}{rms[x(t)]}
    \label{cm-eq}
    \end{split}
\end{equation}

Through simulations we have obtained the CDF of CM and PAPR for all the sequences discussed.
\begin{itemize}
    \item For ZC sequence we have set $ zeroCorrelationZoneConfig =11$, which translates to $N_{CS}=23$ and $L_{RA}=139$. From \eqref{prcapacityzc}, we obtain the preamble capacity of 828 sequences.  
    \item  For $mZC$, defined by \eqref{prcapMZC} we have considered m-seq with cyclic shifts of 2 samples. This results in $N_{m}=70$ different M-sequences, we obtain the preamble capacity of 58788 sequences.
    \item For $aZC$, from \eqref{prcapAZC}, similar to $mZC$ we generate 57960 different sequence.
    \item For $mALL$, in \eqref{eq:mALL}, we generate limited number of sequences 57960 by keeping $l$ constant and varying $\lambda$ from 0 to 137, $w$ from 0 to 69 and $N_{CS}=23$. For different values of $l$ we have same results for PAPR and CM measurements.  
\end{itemize}

  \begin{figure}
    \centering
     \includegraphics[width=0.6\linewidth]{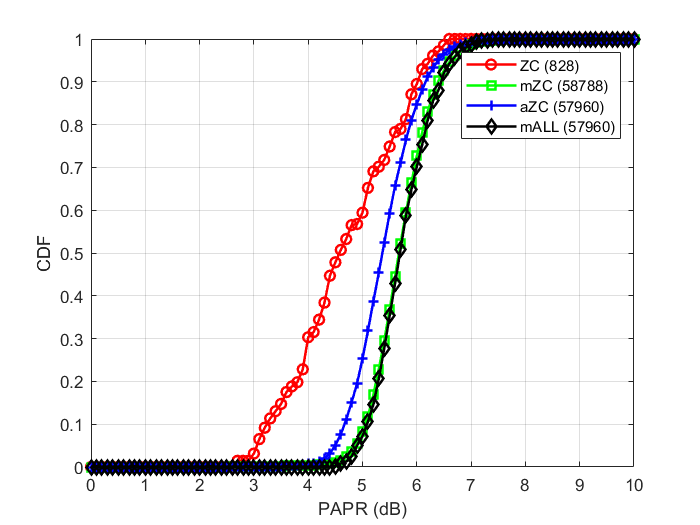}
     \caption{CDF of PAPR for $ZC$, $mZC$, $aZC$, $mALL$  sequences  }
     \label{fig:papr}
     \vspace{-4mm}
 \end{figure}

Note that in above process, we have not considered all possible combinations of the parameters defined in Eq. \eqref{x_c}, \eqref{eq:mZCseq}, \eqref{alltopzseq} and \eqref{eq:mALLcs} due to resource limitations for computations. Considering the length of sequences $L_{RA}=139$, the maximum number of different sequences that can be generated from $ZC$, $mZC$, $aZC$ and $mALL$ sequences using the Preamble capacities defined in Eq. \eqref{prcapacityzc}, \eqref{prcapMZC}, \eqref{prcapAZC} and \eqref{prcapMALL} are summarised in TABLE- \ref{tab:number_of_sequences}. At $N_{CS}=2$, as defined in \cite{TS38.211} for all cases, we can obtain maximum number of different sequences. 

\begin{table}[b]
\vspace{-4mm}
     \centering
      \caption{Maximum number of sequences that can be generated where $L_{RA}=139$ and $N_{CS}=2$ }
     \label{tab:number_of_sequences}
     \begin{tabular}{|p{1.5cm}|p{4cm}|} 
     \hline
     \textbf{Sequences} & \textbf{Maximum Number of sequences}  \\
     \hline
     $ZC$ & $9522$  \\
     \hline
     $mZC$ & $\approx 1.333 \times 10^{6}$    \\ %1333080
     \hline
     $aZC$ & $\approx 1.333 \times 10^{6} $   \\ %1333080
     \hline
     $mALL$ & $\approx  1.853 \times 10^{8}$ \\ %185307711
     \hline
     \end{tabular}
 \end{table}

 We have simulated PAPR and CM measurements using following steps:
 \begin{itemize}
     \item Considering time domain sequences defined in \eqref{zc}, \eqref{setmseq}, \eqref{alltopzseq} and \eqref{eq:mALL}, calculate its FFT of length 139 to convert them into frequency domain.
     \item Map these to 4096 sub-carriers to be generated as OFDM symbols and take IFFT to form time domain signal.
     \item Add the cyclic-prefix (CP) of length 288 which will give us the final time domain signal to be transmitted.
     \item Based on the transmitted time domain signal calculate PAPR and CM using \eqref{papr-eq} and \eqref{cm-eq}. 
 \end{itemize}

 \begin{figure}[t]
    \centering
     \includegraphics[width=0.6\linewidth]{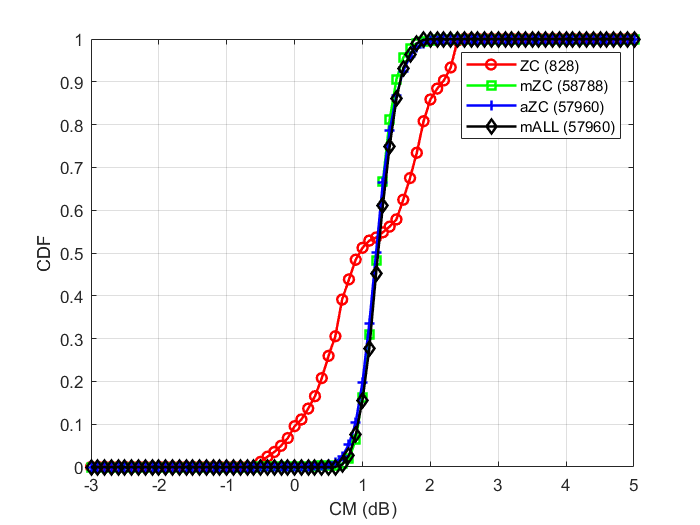}
     \caption{CDF of cubic metric (CM) for $ZC$, $mZC$, $aZC$, $mALL$ sequences}
     \label{fig:cm}
     \vspace{-4mm}
 \end{figure}
 
In Fig. \ref{fig:papr}, we see that the $ZC$ sequence performs better than $mZC$, $aZC$ and $mALL$ sequences. At $99\%$ usage of sequences we have PAPR of 6.6dB, 6.8dB, 7.1dB and 7.05 dB for $ZC, aZC, mZC, mALL$ sequences respectively. This is a very marginal change and can be accommodated when handling other data signals.
In Fig. \ref{fig:cm}, we observe that for CM values between -1dB to 1dB, $ZC$ sequences perform better than $mZC$, $ZC$ and $mALL$. But considering at $50\%$ usage of sequences, all sequences perform equally well at 1.2dB. We see a difference in performance at $99\%$ usage of sequences, where $mZC$ and $mALL$ sequence achieve a CM value of 1.8dB, $aZC$ sequence achieves a CM value of 1.9dB and $ZC$ sequence achieves a CM value of 2.4dB.

From Figs. \ref{fig:papr} and \ref{fig:cm}, it is observed that the proposed $mALL$ sequence has an increased capacity for preambles with a negligible loss in PAPR performance. Also, the CM performance of $mZC$,  $aZC$ and $mALL$ is better than $ZC$ by 0.6dB, 0.5dB and 0.6dB respectively.

\begin{figure*}
     \centering
     \begin{subfigure}[b]{0.45\linewidth}
         \centering
         \includegraphics[width=\linewidth]{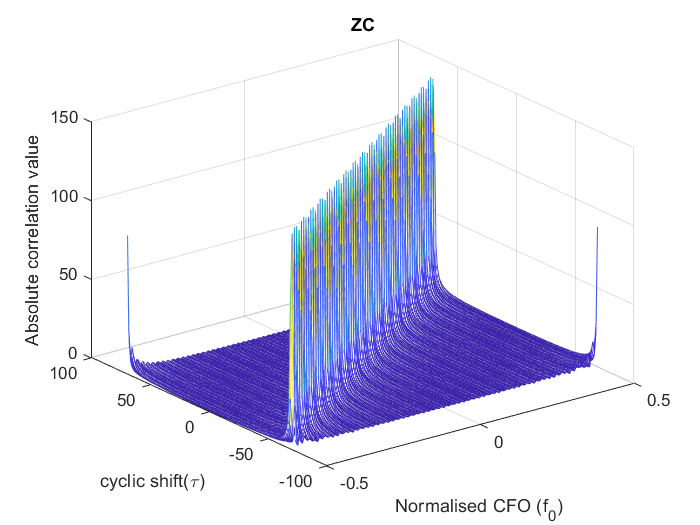}
         \caption{ZC sequence- $x_{2}^{0}$ from Equation. \eqref{x_c}}
         \label{fig:cfo-zc}
     \end{subfigure}
      \hfill
     \begin{subfigure}[b]{0.45\linewidth}
         \centering
         \includegraphics[width=\linewidth]{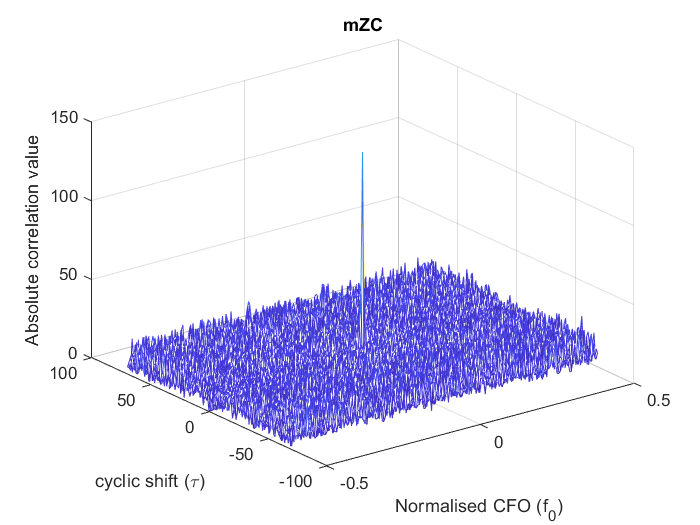}
         \caption{mZC sequence- $z_{1,2,0}$ from Equation. \eqref{eq:mZCseq}}
         \label{fig:cfo-mzc}
     \end{subfigure}
     \hfill
     \begin{subfigure}[b]{0.45\linewidth}
         \centering
         \includegraphics[width=\linewidth]{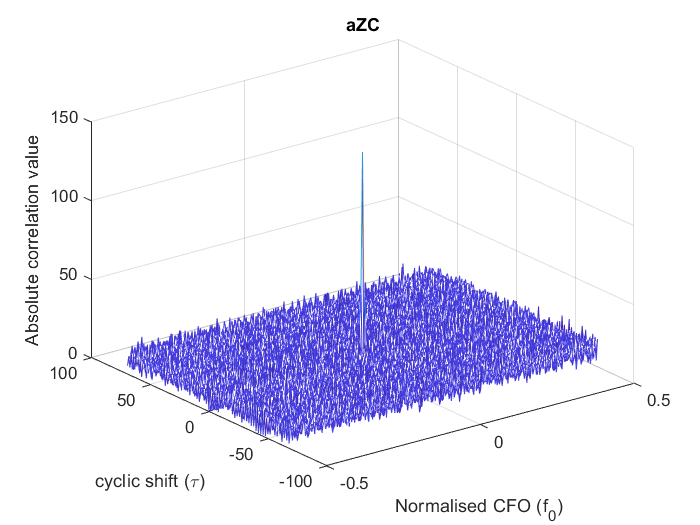}
         \caption{aZC sequence- $z_{1,1,2,1,0}$ from Equation. \eqref{alltopzseq}}
         \label{fig:cfo-azc}
     \end{subfigure}
      \hfill
      \begin{subfigure}[b]{0.45\linewidth}
         \centering
         \includegraphics[width=\linewidth]{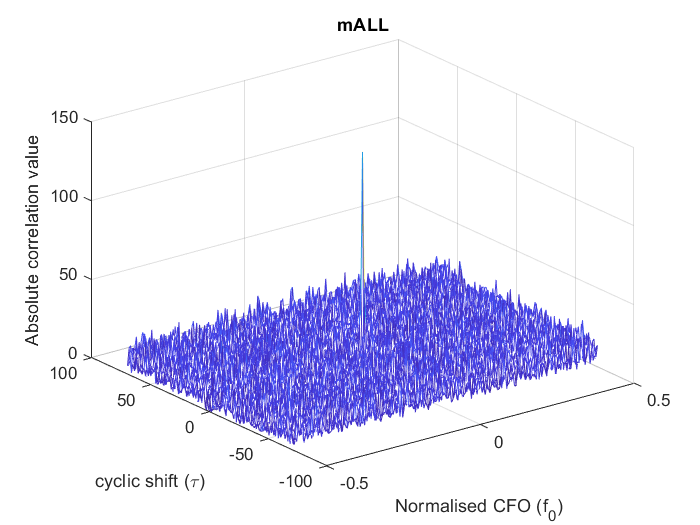}
         \caption{mALL sequence- $z_{1,2,1,0}$ from Equation. \eqref{eq:mALL}}
         \label{fig:cfo-mall}
     \end{subfigure}
        \caption{Periodic Correlation calculated from Equation. \eqref{R_xy} and introducing CFO effect.}
        \label{fig:CFO}
        \vspace{-4mm}
\end{figure*}

\subsection{Effect of Carrier Frequency Offset [CFO]} \label{cfo-effect}
CFO is one of the non-idealities faced by OFDM system. We know that OFDM system is sensitive to timing and frequency synchronisations. CFO occurs when the frequency of local oscillator is not in sync with the carrier frequency. This can occur due to mismatch in transmitter and receiver oscillators and the Doppler effect due to movement of UE. For a received time domain signal $r[n]$, the CFO results in the phase shift of the signal given by \cite{schreiber20185g}:

\begin{equation}
\begin{split}
    r_{f_{0}}[n] =& r[n]*\exp{\left \{ \frac{j2\pi f_{0}n}{L_{RA}} \right \}} \\
    & where \; 0 \leq n \leq L_{RA-1}
    \end{split}
    \label{eq:cfo}
\end{equation}
where $f_{0}$ is the normalised frequency offset due to local oscillator error and Doppler shift. The presence of CFO destroys the orthogonality between sub-carriers and introduces Inter-Carrier Interference (ICI) \cite{ma2003effect}. From \eqref{eq:cfo}, the expression used for calculating the periodic auto-correlation in Eq.\eqref{R_xy} becomes:
\begin{equation}
        R_{x,x}^{f_{0}}[\tau] = \sum_{n=0}^{L-1} x[n] x^{*}[(n+\tau) \; mod \; L_{RA})] e^{\left \{ \frac{j2\pi f_{0}n}{L_{RA}} \right \}}
        \label{eq:cfo-corr}
\end{equation}

The effect of frequency offset is more pronounced for ZC sequences \cite{schreiber20185g}\cite{pitaval2020overcoming}. We make use of Eq. \eqref{eq:cfo-corr} to see the effect of CFO on the auto-correlation properties of $ZC,\;mZC,\;aZC$ and $mALL$ sequences. Fig. \ref{fig:cfo-zc}, shows the auto-correlation result of ZC sequence. We observe that varying the CFO results into high correlation peaks at non-zero values of cyclic shift ($\tau$). This means that in presence of CFO a transmitted ZC sequence with no cyclic shift can be detected as a cyclic shifted version of itself, which is not desirable as it may lead to false-alarm events. Also it can lead to time estimation errors. The $mZC$ and $aZC$ sequences described in \cite{pitaval2020overcoming} are not sensitive to CFO. This can be seen in Fig. \ref{fig:cfo-mzc} and \ref{fig:cfo-azc}, where we observe the peak auto-correlation values at $\tau=0$ and $CFO=0$. Except this, we see low correlation values at all other values ($\tau \neq 0$ and $CFO \neq 0$). This means that we will not detect ambiguous sequences in presence of CFO. For the proposed $mALL$ sequence in Section \ref{proposedmall}, we simulate the auto-correlations in presence of CFO and the results are shown in Fig. \ref{fig:cfo-mall}. We observe similar results compared to  $mZC$ and $aZC$ sequences, where the auto-correlation peak is independent of CFO effect and we face no ambiguity when detecting the transmitted sequence.

\section{Conclusion} \label{conclusion}
In this paper, we have compared ZC sequence detection performance with proposed sequences in literature namely, $mZC$ and $aZC$. We find that the detection performance is very similar for all the three sequences using same detection technique. We propose a new candidate sequence called $mALL$, for preamble generation in RACH procedure. The proposed sequence has a detection performance at par with the $ZC$ sequence which is currently being used in 3GPP standards. The PAPR and CM performance of $mALL$ sequence is compared with sequences such as $ZC,\;mZC$ and $ aZC$, where the CM performance at $99\%$ usage of sequences for the proposed $mALL$ is very similar to $mZC$ and $aZC$ with, $ZC$ sequence performing the worst. PAPR performance at $99\%$ usage of sequence for proposed sequence is comparable to other sequences considered. We also observe that in presence of CFO, the $mZC,\;aZC$ and $mALL$ sequences do not provide ambiguous detection.

The preamble capacity of proposed $mALL$ sequence is two orders of magnitude larger than $mZC,\;aZC$ and four orders of magnitude larger than $ZC$ sequence (TABLE \ref{tab:number_of_sequences}) which is very high, considering no degradation in detection performance, improved CM performance and non-ambiguity in presence of CFO. This increased preamble capacity could help improve the SINR (Signal to Interference and Noise Ratio) and reduction in probability of collisions which remains to be explored.  

\bibliographystyle{IEEEtran}
\bibliography{references}

\end{document}